# Tunable electronic correlation effects in nanotube-light interactions

Yuhei Miyauchi[1,2,3*], Zhengyi Zhang[4], Mitsuhide Takekoshi[5], Yuh Tomio[6], Hidekatsu Suzuura[6], Vasili Perebeinos[7], Vikram V. Deshpande[8,9], Chenguang Lu[10], Stéphane Berciaud[11], Philip Kim[9], James Hone[4], and Tony F. Heinz[1†]

1. Departments of Physics and Electrical Engineering, Columbia University, New York, NY 10027, USA
2. Japan Science and Technology Agency, PRESTO, Saitama 332-0012, Japan
3. Institute of Advanced Energy, Kyoto University, Gokasho, Uji 611-0011, Japan
4. Department of Mechanical Engineering, Columbia University, New York, New York 10027, USA
5. Department of Electrical Engineering, Columbia University, New York, New York 10027, USA
6. Division of Applied Physics, Graduate School of Engineering, Hokkaido University, Sapporo 060-8628, Japan
7. IBM Research Division, T. J. Watson Research Center, Yorktown Heights, New York 10598, USA
8. Department of Physics & Astronomy, University of Utah, Salt Lake City, Utah 84112, USA
9. Department of Physics, Columbia University, New York, New York 10027, USA
10. Department of Applied Physics and Applied Mathematics, Columbia University, New York, New York 10027, USA
11. Institut de Physique et Chimie des Matériaux de Strasbourg and NIE, UMR 7504, Université de Strasbourg and CNRS, 23 rue du Loess, BP43, 67034 Strasbourg Cedex 2, France

* miyauchi@iae.kyoto-u.ac.jp,  † tony.heinz@columbia.edu

**Abstract**

Electronic many-body correlation effects in one-dimensional (1D) systems such as carbon nanotubes have been predicted to modify strongly the nature of photoexcited states. Here we directly probe this effect using broadband elastic light scattering from individual suspended carbon nanotubes under electrostatic gating conditions. We observe significant shifts in optical transition energies, as well as line broadening, as the carrier density is increased. The results demonstrate the differing role of screening of many-body electronic interactions on the macroscopic and microscopic length scales, a feature inherent to quasi-1D systems. Our findings further demonstrate the possibility of electrical tuning of optical transitions and provide a basis for understanding of various optical phenomena in carbon nanotubes and other quasi-1D systems in the presence of charge carrier doping.

Quantum many-body correlation effects in photon-matter interactions have long been a subject of central importance in condensed-matter physics. Understanding of these effects in quasi-1D systems, where their influence is generally far more pronounced than in bulk materials[1, 2], has been a focus of particular attention. In this context, single-walled carbon nanotubes (hereafter referred to as nanotubes) provide the ultimate quasi-1D quantum wires. They have diameters on the order of one nanometer and shells of single-atom thickness. Photoexcited electrons and holes in semiconducting nanotubes form stable two-particle excitonic bound states through their mutual Coulomb attraction, with binding energies reaching hundreds of meV[2-9]. Excitons in nanotubes are therefore highly stable even at room temperature, unlike the case for conventional semiconductor quantum wires. Additionally, one can tune the carrier density and Fermi energy in nanotubes widely by means of electrostatic gating[10-14]. Nanotubes thus offer an ideal platform for the fundamental studies of the many-body interactions and their impact on the optical spectra of photoexcited quasi-1D systems. Knowledge of the underlying optical response of nanotubes under carrier doping is also of great importance for the development of electrically tunable optoelectronic devices[13] that operate at room temperature and on the nanometer length scale. Despite intensive effort[3-30], however, complete understanding of such phenomena has been hindered by the difficulty of obtaining optical spectra of isolated 1D nanostructures over a wide range of free carrier densities.

In this Letter, we demonstrate the direct characterization of the optical transitions as a function of carrier density using a spectroscopic technique capable of accessing individual nanotubes. The approach provides optimal conditions for the identification of many-body effects on optical transitions and for detailed examination of the underlying electron-electron, electron-hole, exciton-electron (hole), and exciton-photon interactions. Our measurements rely on broadband elastic (Rayleigh) light scattering spectroscopy[23, 31] from individual, suspended carbon nanotubes. The experimental configuration allows for measurement of scattering spectra under conditions of variable electrostatic gating, which provides precise control of the injected charge density (See Fig. 1a for a schematic representation of the measurement). We observed peak shifts and broadening of exciton resonances with increasing carrier density in semiconducting nanotubes. The measured red shift of the exciton transition energies with doping indicates a reduction of strength of the Coulomb interactions, leading to a decrease in the quasi-particle band gap, partially offset by a decrease in the exciton binding energy. The magnitude of the spectral shifts with doping can be understood within the context of a length-scale dependent dielectric screening inherent to quasi-1D systems, on the basis of theoretical calculation of excitons in carrier-doped nanotubes. The width of the excitonic peak is found to increase nearly linearly with injected carrier density. This behavior is well reproduced by calculations of intervalley scattering processes involving the injected charges. Our findings provide fundamental information on the optical transitions and the role of many-body interactions in defining their characteristics in carrier-doped carbon nanotubes. The results will serve a basis for the interpretation of various optical phenomena sensitive to the optical resonances under the doped conditions. The large demonstrated tunability of exciton resonance in these systems also provides the basis for the development of novel optoelectronics devices, such as gate-tunable photon detectors, modulators and metamaterials.

We examined individual single-walled carbon nanotubes prepared by ambient chemical vapor deposition methods [32]. The nanotubes were grown directly over open slits of 20-30 μm width cut in $SiO_2$/Si substrates as shown in Fig.1a. The electrical contact to the nanotubes needed for gating experiments was obtained using electrodes (1 nm of tungsten and 15 nm of platinum) fabricated on both sides of the slit. The gate bias for electrostatic doping was applied to the silicon substrate. The gate electric field from the sidewalls of the slit served to tune the Fermi energy and carrier density in the nanotube (Fig.1b). The gate efficiency of the device was determined using the variation of Raman scattering spectra from a chiral metallic nanotube under gating conditions[32]. We confirmed experimentally that the inhomogeneity of the gating efficiency within a region of ±7.5 μm of center of a nanotube was negligible by comparing the Rayleigh spectra taken at the different positions along the nanotube under the gating conditions.

Broadband light generated by a super-continuum source (with integrated power of about 0.4 mW over the wavelength range of ~450-1100 nm, for Rayleigh spectroscopy) or monochromatic light from a laser (with power of 0.37 mW, for Raman spectroscopy) was focused at the center of an individual suspended nanotube with spot size of 2-3 μm and polarization along the nanotube axis. The scattered light was collected by a spectrometer through a confocal arrangement[31]. Our initial characterization of an individual nanotube is obtained by collecting both elastic (Rayleigh) and inelastic (Raman) scattering spectra. Representative spectra are shown in Figs. 1c and 1d for a near zigzag semiconducting nanotube of about 2-2.1 nm in diameter, tentatively assigned to a (26, 0) species [31-33]. Over the available Rayleigh spectral range, we observe two distinct peaks. The features correspond, respectively, to the third and fourth subband exciton resonances ($S_{33}$ and $S_{44}$) in the near zigzag semiconducting nanotube. All measurements were conducted at room temperature under a nitrogen gas environment to prevent any possible modification from ambient air, for example, through $H_2O$ or $O_2$ adsorption.

Under the application of a gate bias voltage $V_G$ to inject charge into the suspended semiconducting nanotube, we observed clear shifts in the peak positions of the $S_{33}$ and $S_{44}$ transitions, as well as broadening of the features with increasing $|V_G|$ (Fig. 2a). The spectra are analyzed by a fitting procedure using an excitonic description of the nanotube optical response[23]. For Rayleigh scattering of a one-dimensional system, the measured scattering cross-section $\sigma \propto \omega^3 |\chi(\omega)|^2$, where $\omega$ is the optical frequency and $\chi(\omega)$ is the nanotube optical susceptibility. We model the nanotube excitonic response by a Lorentzian line shape of the form $\chi = \chi_b + f[(\omega_0 - \omega) - i\Gamma/2]^{-1}$. Here $\chi_b$ is the (frequency-independent) background susceptibility arising from the non-resonant transitions, $f$ is a parameter proportional to the exciton oscillator strength, $\omega_0$ is the resonance frequency, and $\Gamma$ is the line width of the transition. This fitting procedure has been successfully applied for higher-order Rayleigh scattering peaks of metallic and semiconducting nanotubes[23].

This model generates excellent fits to the experimental scattering spectra (Fig. 2b). For clarity, we discuss the results for negative gate bias voltages $V_G$, corresponding to hole doping; the results for electron doping are similar[32]. We stress that the $S_{33}$ and $S_{44}$ bands are not directly filled by the injected electrons or holes within the observed range of $|V_G|$[32]. These carriers reside only within the lowest energy bands (Fig. 1b). Thus, simple Pauli blocking effects influence only the $S_{11}$ transition. By observing the higher-order $S_{33}$ and $S_{44}$ transitions, we directly probe the modification of the *many-body interactions* induced by the free-carrier density.

The experimentally determined shift in the exciton transition energies of the near zigzag semiconducting nanotube as a function of the injected carrier density $\rho$ is presented in Fig. 3a. In order to focus on modifications of the electronic many-body interactions under the gating conditions, we plot the average value of the shifts of $S_{33}$ and $S_{44}$ peaks, $\Delta E_{av}$. This procedure eliminates any effects from electrostatically-induced strain[32], which causes the two peaks to shift in opposite directions, but does not change their average position[34]. We found that the observed red shift of the exciton energy of the semiconducting nanotube with charge density $\rho$ is reproduced using a power-law function of the form $\Delta E_{av} = A\rho^\alpha$ (dotted line in Fig. 3a), with an exponent of $\alpha = 0.6$ and a prefactor of $A = 106$ meV·nm$^{0.6}$.

In order to understand the origin of the observed peak shifts, we conducted a theoretical study of excitons within the $k \cdot p$ approximation[2, 26]. The $S_{33}$ and $S_{44}$ exciton energies dependent on carrier density are predicted within a $k \cdot p$ scheme by solving the Bethe-Salpeter (BS) equation[32]. In the calculation, exciton states are determined by a circumferential length $L$ (= $\pi d$) of the nanotube, a characteristic kinetic energy $2\pi\gamma/L$ [$\gamma = (\sqrt{3}/2)a\gamma_0$ with nearest-neighbor hopping integral $\gamma_0 = 2.7$ eV and a lattice constant $a = 2.46$ Å of graphene], and a characteristic

Coulomb energy $e^2/\kappa L$, where $\kappa$ is an effective dielectric constant accounting for the effects of polarization of surrounding materials and electrons far from the Fermi level. The dimensionless Coulomb parameter $v = (e^2/\kappa)/(2\pi\gamma)$ represents the Coulomb interaction $e^2/\kappa L$ scaled by the characteristic kinetic energy. In the calculations we use, $v = 0.16$, corresponding to $\kappa = 2.5$. In our $k \cdot p$ treatment, we consider intravalley Coulomb interactions near the $K$ ($K'$) valley of the graphene Brillouin zone. We calculate the screened Coulomb interaction that depends on the carrier density and the relevant length scale of the interactions within the static random phase approximation[32]. All the calculations were conducted on a (26, 0) semiconducting nanotube for comparison with the experimental results.

Figure 3b displays the calculated shift for the $S_{33}$ energy, $\Delta E_{33}$ (red dot-dashed curve), for the $S_{44}$ exciton energy, $\Delta E_{44}$ (red doted curve), and for their average (red solid curve) as a function of injected carrier (hole) density. The experimental data $\Delta E_{av}$ are plotted together for comparison (blue circles). The trend of the experimental energy shift is well reproduced for typical Coulomb parameters[32]. The calculation provides a clear perspective on the physical origin of the shift in exciton energy induced by free carriers. As shown in Fig. 3c, the exciton transition energy $E_{ex}$ is determined by

$$E_{ex} = E_{sp} + \Sigma - E_b, \qquad (1)$$

where $E_{sp}$ is a single particle gap between the relevant subbands, $\Sigma$ is the self-energy correction for the quasi-particle band gap from electron exchange correlations, and $E_b$ is the exciton binding energy from the attractive electron-hole interactions. (Here we consider $\Sigma$ and $E_b$ to be positive quantities.) Therefore, the net many-body correction energy $E_{m-b}$ to the single particle gap is $E_{m-b} = \Sigma - E_b$, and the exciton energy shift due to modified many-body interactions originates from $\Delta E_{m-b}$ (=$\Delta(\Sigma - E_b)$), as shown in Fig. 3c.

The calculated changes in $E_b$ and $\Sigma$ ($\Delta E_b$ and $\Delta\Sigma$, averaged for $S_{33}$ and $S_{44}$) are plotted in Fig. 3b to show separately the contributions of each component to the net exciton energy shift $\Delta E_{m-b}$.

We find that $\Delta E_b$ and $\Delta\Sigma$ cancel out to a significant degree, but that $|\Delta\Sigma|$ is always larger than $|\Delta E_b|$. Thus a red shift of the exciton energy $E_{ex}$ is always predicted ($\Delta E_{m-b} < 0$), in accordance with experiment. The key to understanding the mechanism of the net red shift lies in the dependence of the Coulomb interaction on the length scale[6]. The value of $\Sigma$ in Eq. 1 consists of the contributions both from long-range 1D Coulomb interactions $\Sigma_l$ (occurring on a length scale greater than the nanotube circumference) and from 2D-like short-range interactions $\Sigma_s$ (occurring on a length scale less than the circumference). Given the spatial extent of the exciton, the exciton binding energy $E_b$, on the other hand, is dominated by long-range 1D Coulomb interactions[6, 27]. The changes in $\Sigma_l$ and $E_b$ relevant to the 1D Coulomb interactions are large, but cancel one another to a high degree of accuracy, whether in the neutral or doped system. Therefore, the shift in the exciton energy $\Delta E_{m-b}$ is almost completely dominated by the change in $\Sigma_s$, namely, $\Delta E_{m-b} \approx \Delta\Sigma_s$. We note that our calculations have been carried out in the limit of static screen. Since the dynamical effects have been predicted to weaken screening of 1D interactions [25], this may lead to an underestimate of values $E_b$ and $\Sigma$. The discussion above is, however, robust as long as the cancellation of the long-range 1D interactions is maintained.

Let us therefore consider how $\Sigma_s$ is reduced with doping using our experimental observations. The exciton transition energies in our near zigzag semiconducting nanotube show a red shift of about 60 meV or 3% of the transition energy at a charge density of $\rho = 0.4$ $e$/nm. Our calculation of the excitons in a (26, 0) nanotube suggests that $E_{m-b}$ ($\approx \Sigma_s$) accounts for about 20% of the net transition energy in the undoped nanotube. Thus, the 3% red shift of the exciton energy corresponds to ~15% decrease in $\Sigma_s$, *i.e.*, the short-range interaction strength is reduced to 85% of its intrinsic strength for $\rho = 0.4$ $e$/nm. This tunability of the short-range interaction strength for a wide range of free carrier density, together with the nearly perfect cancellation of $\Sigma_l$ and $E_b$ by 1D long-range interactions, makes fine control of the higher sub-band exciton resonance energy possible without relying on the Pauli-blocking effect in carbon nanotubes.

We now consider another important experimental observation, namely, the change in the line width of the excitonic transitions with doping. Figure 4a shows the carrier density dependence of the line width of the $S_{33}$ and $S_{44}$ transitions in the near zigzag semiconducting nanotube. We find that the line widths increase significantly with carrier density $\rho$. A similar effect has recently been reported for higher energy $S_{44}$ and $S_{55}$ excitons in individual semiconducting nanotubes on substrates using an optical microscopy technique [35]. The line width of an excitonic transition reflects the rate of dephasing of the exciton coherence, and the inferred exciton dephasing times with doping are shown on the right axis of Fig. 4a. The observed line width change exhibit a nearly linear increase with increasing $\rho$, except for the step-like increase of about 10 meV at low $\rho$. For densities $\rho > 0.05$ $e$/nm, the experimental line width $\Gamma(\rho)$ can be fit with a linear function of carrier density as $\Gamma_{ii}(\rho) = \Gamma_{ii0} + \beta_{ii}\rho$, where $\Gamma_{ii0}$ and $\beta_{ii}$ are constants. The coefficients $\beta_{33} = 145$ meV·nm and $\beta_{44} = 114$ meV·nm characterize the additional contribution of doping to the dephasing of the $S_{33}$ and $S_{44}$ excitons.

Our theoretical treatment for the peak shift within the treatment of the BS equation also predicts the exciton line width as the imaginary part of the complex exciton energy[32]. We found, however, that the additional line width is nearly constant for finite carrier densities of $\rho > 0.05$ $e$/nm, and the linear increase of the line width is not reproduced. To address this discrepancy with theory, we explored possible contributions of the *intervalley* (*K-K'*) interactions to the line width that were omitted within the treatment of the *k·p* approximation[32]. To calculate the decay rates and line broadening of the excitonic transitions we consider the process in which a photoexcited electron-hole pair in the third (or fourth) subband scatters with injected electrons or holes in the first subband (Fig. 4b). We find that there are available scattering pathways for photoexcited electron and hole in an electron or hole doped nanotube that satisfy energy-momentum conservation for the intervalley scattering as shown in Fig. 4b. The corresponding rates are evaluated within the second order perturbation theory, treating the intervalley scattering pathways shown in Fig. 4b with the same Coulomb parameters as were used for the calculation of the peak shifts. As expected, a peak shift also arises from these intervalley scattering processes. The calculated shift at $\rho = 0.42$ $e$/nm (~2.6 meV) is, however, far smaller than the experimental shift (~60 meV) in Fig. 3. Hence, we can safely neglect the effect of the intervalley processes in describing doping-induced shifts of the excitonic transition energies.

Fig. 4c shows the calculated line width broadening (the sum of decay rate of photoexcited electron and hole $\Gamma_e + \Gamma_h$) as a function of carrier density $\rho$, plotted together with the same experimental data shown in Fig. 4a. In good agreement with experiment, we predict nearly linear increases of the line width of $S_{33}$ and $S_{44}$ transitions. The calculation also shows that the number of decay channels for photo-excited holes in the $S_{44}$ band is smaller than that for $S_{33}$. This explains the counter-intuitive experimental result of a slightly smaller doping-induced broadening of $S_{44}$ width than that of $S_{33}$[32]. We therefore attribute the nearly linear broadening of the line width with doping to intervalley scattering processes involving the injected charges (Fig. 4b).

Finally, we comment briefly on some practical implications of this study. The observed tunability of the exciton resonance frequency by a full line width for carrier levels on the order of only 0.5 $e$/nm could be useful for applications as nano-sized, tunable photon detectors or modulators. Furthermore, such tunability of the optical responses may offer the possibility of constructing tunable metamaterials by integration of individual nanotubes in other structures. Our findings are also important for the interpretation of various experimental results related to the optical resonance of nanotubes. In particular, our results imply that one needs to exercise considerable care in making chirality assignment of nanotubes on substrates, where unintentionally doping effects may lead to a significant shift in the measured transitions energies. More broadly, the strong many body effects in these materials imply that their optical response will be significantly modified by carrier density, even in the absence of Pauli blocking effects.

In conclusion, we have demonstrated that the higher-order excitonic transitions in semiconducting carbon nanotubes exhibit significant red shifts and broadening with increasing carrier doping. These features do not arise from simple Pauli blocking, since no charge is injected into the relevant bands, but rather reflect the role of many-body electronic interactions. Our theoretical investigations show that the decrease in many-body contributions to the excited-state energy is dominated by intravalley interactions with a quasi-2D character. The nearly linear peak broadening with increased carrier density, on the other hand, has a different physical origin: It can be explained on the basis of intervalley scattering processes involving the injected carriers. These results provide a rigorous basis for the interpretation of changes in the optical response induced by carrier doping in carbon nanotubes and other quasi-1D systems; the results also suggest new possibilities for gate-tunable optoelectronic devices.

**Acknowledgements**

This research was supported by a Grant-in-Aid for Scientific Research Grants 08J03712 and 24681031 from Japan Society for the Promotion of Science (JSPS) and by Precursory Research for Embryonic Science and Technology (PRESTO) Grant 3538 from Japan Science and Technology Agency (JST) for Y.M. The spectroscopic studies were supported by the US National Science Foundations through grant DMR-1106225; device fabrication was supported by the US Department of Energy, Office of Basic Energy Sciences through grant DE-SC0001085 for the Columbia Energy Frontier Research Center, and by the Honda Research Institute. The authors would like to thank P. Avouris and M. Freitag for fruitful discussions.

**References**

[1] T. Ogawa, and T. Takagahara, Phys. Rev. B **44**, 8138 (1991).
[2] T. Ando, J. Phys. Soc. Jpn. **66**, 1066 (1997).
[3] F. Wang, G. Dukovic, L. E. Brus, and T. F. Heinz, Science **308**, 838 (2005).
[4] J. Maultzsch, R. Pomraenke, S. Reich, E. Chang, D. Prezzi, A. Ruini, E. Molinari, M. S. Strano, C. Thomsen, and C. Lienau, Phys. Rev. B **72**, 241402(R) (2005).
[5] Y.-Z. Ma, L. Valkunas, S. M. Bachilo, and G. R. Fleming, J. Phys. Chem. B **109**, 15671 (2005).
[6] C. L. Kane, and E. J. Mele, Phys. Rev. Lett. **93**, 197402 (2004).
[7] C. D. Spataru, S. Ismail-Beigi, L. X. Benedict, and S. G. Louie, Phys. Rev. Lett. **92**, 077402 (2004).
[8] V. Perebeinos, J. Tersoff, and P. Avouris, Phys. Rev. Lett. **92**, 257402 (2004).
[9] H. Zhao, and S. Mazumdar, Phys. Rev. Lett. **93**, 157402 (2004).
[10] L. Marty, E. Adam, L. Albert, R. Doyon, D. Ménard, and R. Martel, Phys. Rev. Lett. **96**, 136803 (2006).
[11] M. Steiner, M. Freitag, V. Perebeinos, A. Naumov, J. P. Small, A. A. Bol, and P. Avouris, Nano Lett. **9**, 3477 (2009).
[12] S. Yasukochi, T. Murai, S. Moritsubo, T. Shimada, S. Chiashi, S. Maruyama, and Y. K. Kato, Phys. Rev. B **84**, 121409(R) (2011).
[13] P. Avouris, M. Freitag, and V. Perebeinos, Nature Photon. **2**, 341 (2008).
[14] A. W. Bushmaker, V. V. Deshpande, S. Hsieh, M. W. Bockrath, and S. B. Cronin, Phys. Rev. Lett. **103**, 067401 (2009).
[15] T. Hertel, R. Fasel, and G. Moos, Appl. Phys. A **75**, 449 (2002).
[16] J. Kono, G. N. Ostojic, S. Zaric, M. S. Strano, V. C. Moore, J. Shaver, R. H. Hauge, and R. E. Smalley, Appl. Phys. A **78**, 1093 (2004).
[17] A. Jorio, C. Fantini, M. A. Pimenta, R. B. Capaz, G. G. Samsonidze, G. Dresselhaus, M. S. Dresselhaus, J. Jiang, N. Kobayashi, A. Grüneis, and R. Saito, Phys. Rev. B **71**, 075401 (2005).
[18] L. Cognet, D. A. Tsyboulski, J.-D. R. Rocha, C. D. Doyle, J. M. Tour, and R. B. Weisman, Science **316**, 1465 (2007).
[19] A. G. Walsh, A. N. Vamivakas, Y. Yin, S. B. Cronin, M. S. Ünlü, B. B. Goldberg, and A. K. Swan, Nano Lett. **7**, 1485 (2007).
[20] K. Sato, R. Saito, J. Jiang, G. Dresselhaus, and M. S. Dresselhaus, Phys. Rev. B **76**, 195446 (2007).
[21] J. Lefebvre, and P. Finnie, Nano Lett. **8**, 1890 (2008).
[22] L. Lüer, S. Hoseinkhani, D. Polli, J. Crochet, T. Hertel, and G. Lanzani, Nature Phys. **5**, 54 (2009).
[23] S. Berciaud, C. Voisin, H. Yan, B. Chandra, R. Caldwell, Y. Shan, L. E. Brus, J. Hone, and T. F. Heinz, Phys. Rev. B **81**, 041414(R) (2009).
[24] J. Deslippe, M. Dipoppa, D. Prendergast, M. V. O. Moutinho, R. B. Capaz, and S. G. Louie, Nano Lett. **9**, 1330 (2009).
[25] C. D. Spataru, and F. Léonard Phys. Rev. Lett. **104**, 177402 (2010).
[26] Y. Tomio, H. Suzuura, and T. Ando, Phys. Rev. B **85**, 085411 (2012).
[27] K. Liu, J. Deslippe, F. Xiao, R. B. Capaz, X. Hong, S. Aloni, A. Zettl, W. Wang, X. Bai, S. G. Louie, E. Wang, and F. Wang, Nature Nanotech. **7**, 325 (2012).
[28] S. Mouri, and K. Matsuda, J. Appl. Phys. **111**, 094309 (2012).
[29] S. Konabe, K. Matsuda, and S. Okada, Phys. Rev. Lett. **109**, 187403 (2012).
[30] Y. Kimoto, M. Okano, and Y. Kanemitsu, Phys. Rev. B **87**, 195416 (2013).
[31] M. Y. Sfeir, T. Beetz, F. Wang, L. Huang, X. M. H. Huang, M. Huang, J. Hone, S. O'Brien, J. A. Misewich, T. F. Heinz, L. Wu, Y. Zhu, and L. E. Brus, Science **312**, 554 (2006).
[32] See supplementary material at [URL will be inserted] for a detailed discussion.

[33] S. Berciaud, V. V. Deshpande, R. Caldwell, Y. Miyauchi, C. Voisin, P. Kim, J. Hone, and T. F. Heinz, phys. stat. sol. b **249**, 2436 (2012).

[34] M. Huang, Y. Wu, B. Chandra, H. Yan, Y. Shan, T. F. Heinz, and J. Hone, Phys. Rev. Lett. **100**, 136803 (2008).

[35] K. Liu, X. Hong, Q. Zhou, C. Jin, J. Li, W. Zhou, J. Liu, E. Wang, A. Zettl, and F. Wang, arXiv:1307.5353 [cond-mat.mes-hall]  (2013).

**Figure Captions**

**Figure 1** (a) Schematic of our individual nanotube device. Broadband white light is focused onto the nanotube suspended across the gap. The inset shows a microscope image of a substrate with the slit and electrodes used in this study. (b) Schematic diagram of the band structure of a semiconducting nanotube under electrostatic gating conditions (hole doping). The dotted horizontal line indicates the Fermi energy $E_F$. Blue filled and empty circles represent photoexcited electrons and holes associated with the $S_{33}$ and $S_{44}$ transitions in a semiconducting nanotube. $v_i$ and $c_i$ are $i$-th valence and conduction subbands, respectively. (c) Elastic (Rayleigh) and (d) inelastic (Raman) light scattering spectra of a single undoped nanotube. The Raman spectra were recorded using a laser photon energy of 1.96 eV. The inset shows the spectrum around the radial breathing mode (RBM). We present all the Rayleigh spectra corrected for the $\omega^3$ scattering efficiency factor to reflect directly the optical susceptibility.

**Figure 2** (a) Rayleigh scattering spectra for the observed semiconducting nanotube for gating voltages from $V_G$= 0V to -25V. Similar spectral changes were observed for positive gating voltages[32]. (b) Rayleigh scattering spectra of the semiconducting nanotube at $V_G$= 0V (black solid circles) and $V_G$=-25V (red solid circles). The solid green curves are fits based on an excitonic model of the optical susceptibility described in the main text.

**Figure 3** (a) Dependence of the exciton resonance energy shifts of the semiconducting nanotube on the carrier (hole) density. We plot the average peak shifts of $S_{33}$ and $S_{44}$ transitions to eliminate possible effects of strain[32]. The dotted curve indicates a power-law function: $\Delta E_{av} = A\rho^{\alpha}$, with parameters $\alpha = 0.6$ and $A = 106$ meV·nm$^{0.6}$. (b) Calculated $S_{33}$ and $S_{44}$ shifts of the exciton transition energies $\Delta E_{33}$ and $\Delta E_{44}$ (red dot-dashed and doted curves) and their average (red solid curve) for a (26, 0) nanotube as a function of carrier density $\rho$[32]. The calculated changes in $E_b$ and $\Sigma$ ($\Delta E_b$ and $\Delta\Sigma$) (blue and green curves, averaged for $S_{33}$ and $S_{44}$ excitons), and the experimental data $\Delta E_{av}$ shown in Fig. 4a (blue circles) are plotted together for comparison. At $\rho=0.4$ $e$/nm, $E_b$ and $\Sigma$ for $S_{33}$ are reduced to be 21% (50 meV) and 51% (349 meV) of their original values for the undoped nanotube, respectively. (c) Schematic diagram of the physical mechanisms for the red shift in the transition energy with doping. The exciton energy $E_{ex}$ (red horizontal line) decreases with increasing doping according to $\Delta E_{m-b} = \Delta(\Sigma - E_b)$, as discussed in the text.

**Figure 4** (a) Dependence of the $S_{33}$ and $S_{44}$ exciton resonance line widths of the semiconducting nanotube on carrier density. The dotted lines are fits of the function $\Gamma_{ii}(\rho) = \Gamma_{ii0} + \beta_{ii}\rho$ to the line widths for the $S_{ii}$ exciton, with $\beta_{33} = 145$ meV·nm and $\beta_{44} = 114$ meV·nm, respectively. The right axis indicates the dephasing time of excitons corresponding to the line width. (b) Schematic representation of the possible electron (hole) scattering pathways (pairs of arrows with the same colors) contributing to the observed increase in line width. Only the case for an electron-doped nanotube is shown. (The same scattering rate is obtained for both electron- and hole-doped nanotubes because of the electron-hole symmetry in our theoretical treatment.) For $S_{33}$ ($S_{44}$), only the pathways where a photoexcited electron and a hole in $K$ ($K'$) valley are involved are shown. The opposite pathways through the $K'$ ($K$) valley are not shown, but are taken into account in the calculation. (c) Calculated decay rates (red and green solid lines for $S_{33}$ and $S_{44}$, respectively) of a photoexcited electron-hole pair in a (26, 0) nanotube due to the free carrier doping[32]. The same Coulomb parameters ($\gamma_0 = 2.7$ eV and $\kappa = 2.5$) are used as for calculation of the shifts in exciton transition energies shown in Fig. 3(b). The calculated data is plotted together with the experimental data (opaque symbols) shown in (a) with 65 meV offset between the left (for calculated results) and right (for experimental data) vertical axes to account for decay channels of the undoped nanotube.

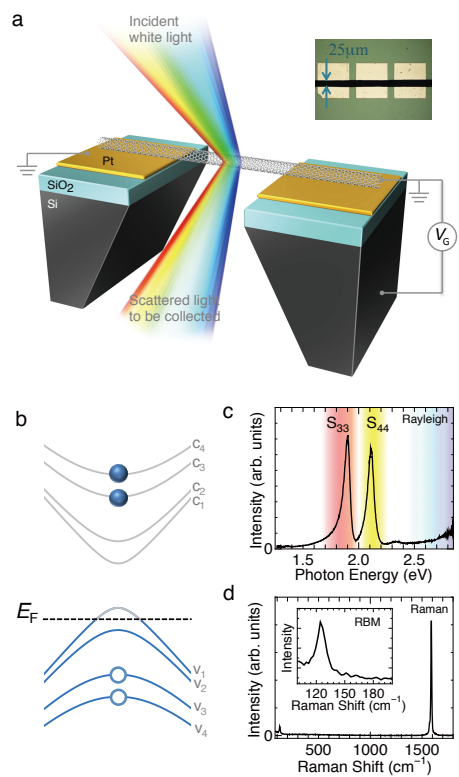

Figure 1. Y. Miyauchi *et al.*

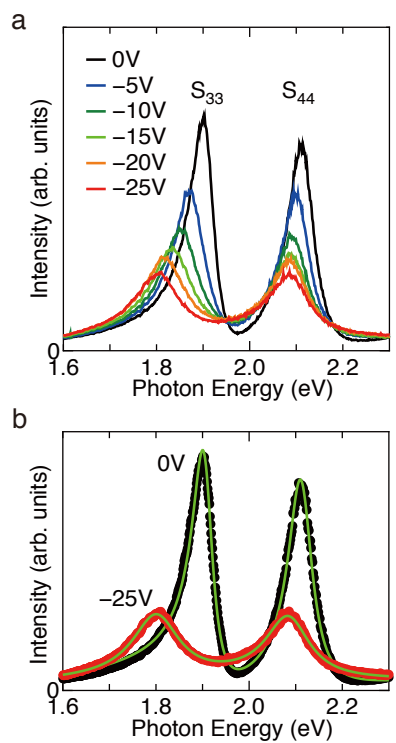

Figure 2. Y. Miyauchi *et al.*

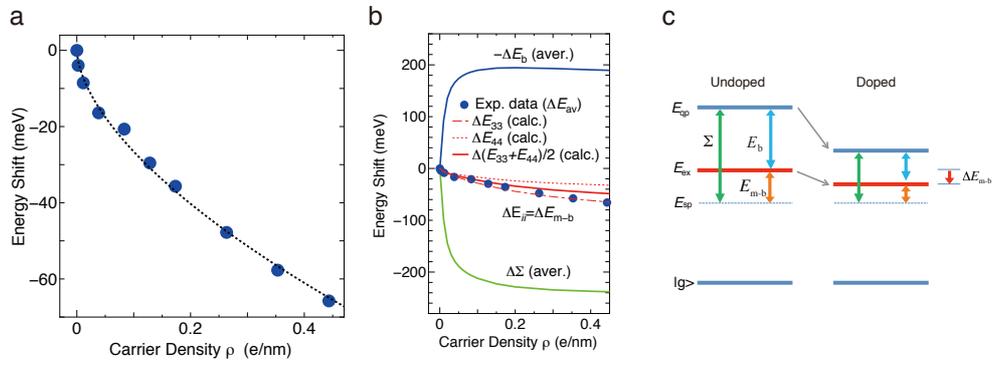

Figure 3. Y. Miyauchi *et al*.

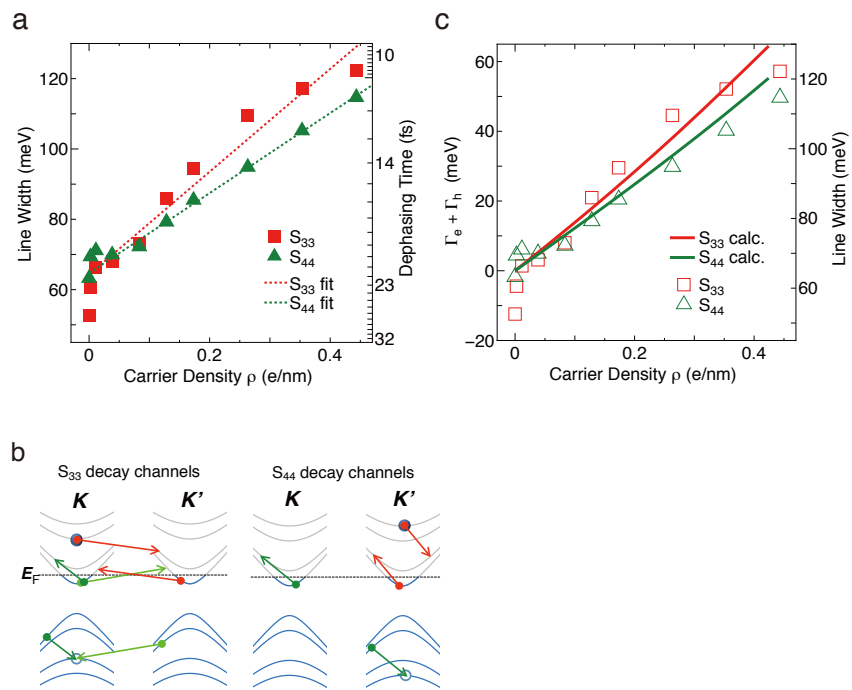

Figure 4. Y. Miyauchi *et al.*

Supplemental Material for

# "Tunable electronic correlation effects in nanotube-light interactions"

Yuhei Miyauchi[1,2,3*], Zhengyi Zhang[4], Mitsuhide Takekoshi[5], Yuh Tomio[6], Hidekatsu Suzuura[6], Vasili Perebeinos[7], Vikram V. Deshpande[8,9], Chenguang Lu[10], Stéphane Berciaud[11], Philip Kim[9], James Hone[4], and Tony F. Heinz[1†]

1. Departments of Physics and Electrical Engineering, Columbia University, New York, NY 10027, USA
2. Japan Science and Technology Agency, PRESTO, Saitama 332-0012, Japan
3. Institute of Advanced Energy, Kyoto University, Gokasho, Uji 611-0011, Japan
4. Department of Mechanical Engineering, Columbia University, New York, New York 10027, USA
5. Department of Electrical Engineering, Columbia University, New York, New York 10027, USA
6. Division of Applied Physics, Graduate School of Engineering, Hokkaido University, Sapporo 060-8628, Japan
7. IBM Research Division, T. J. Watson Research Center, Yorktown Heights, New York 10598, USA
8. Department of Physics & Astronomy, University of Utah, Salt Lake City, Utah 84112, USA
9. Department of Physics, Columbia University, New York, New York 10027, USA
10. Department of Applied Physics and Applied Mathematics, Columbia University, New York, New York 10027, USA
11. Institut de Physique et Chimie des Matériaux de Strasbourg and NIE, UMR 7504, Université de Strasbourg and CNRS, 23 rue du Loess, BP43, 67034 Strasbourg Cedex 2, France

\* miyauchi@iae.kyoto-u.ac.jp,   † tony.heinz@columbia.edu

**S1. Sample preparation and device fabrication**

We examined individual single-walled carbon nanotubes prepared by ambient chemical vapor deposition methods using a modified fast-heating process[S1]. The nanotubes were grown directly over open slits of 20-30 μm width cut in $SiO_2$/Si substrates. The electrical contact to the nanotubes needed for gating experiments was obtained using electrodes (1 nm of tungsten and 15 nm of platinum) fabricated on both sides of the slit. The chiral indices of each nanotube spanning the slit were assigned by simultaneous Rayleigh and Raman spectroscopy[S2, 3]. For a nanotube employed in the measurements described in the main text, we observed two distinct peaks at 1.9 and 2.1 eV in the Rayleigh scattering spectrum, and G-mode (at 1592 cm$^{-1}$) and radial breathing mode (at 125 cm$^{-1}$) features in the Raman spectrum as shown in Fig. 1. On the basis of these observations, the nanotube is assigned to a near zigzag semiconducting nanotube of 2.0-2.1 nm in diameter, belonging to the family with chiral indices (*n*, *m*) satisfying (*n* - *m*) mod 3 = 2. The possible assignment of the nanotube structure is a zigzag (26, 0), or possibly neighboring near zigzag species such as (25, 2)[S2, 3]. The narrow G-mode feature at 1592 cm$^{-1}$ in the Raman spectrum indicating a dominant contribution of zone center longitudinal optical photon is also consistent with the assignment to near zigzag semiconducting species[S3]. Unwanted nanotubes were removed by intense laser irradiation to yield a clean experimental geometry for the gating measurements.

**S2. Evaluation of the gate efficiency and carrier density**

The gate efficiency of our nanotube devices determines the relationship between the applied voltage and the physically important induced charge density. We evaluated this quantity using Raman scattering spectroscopy of a chiral metallic nanotube suspended over the same slit where the semiconducting nanotube employed in the measurements described in the main text was located. The line width of the G⁻ Raman mode of chiral metallic carbon nanotubes is known to change systematically as a function of the charge density because of the dependence of the phonon lifetime[S4, 5] on the position of the Fermi energy $E_F$.

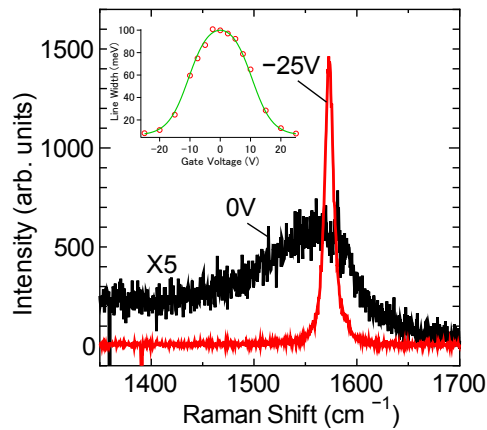

**Figure S1.** Raman G$^-$ mode spectra for gate voltages of $V_G$=0 (black curve) and −25V (red curve). The inset displays the measured width of this mode (red circles) as a function of the gate voltage $V_G$. The green curve is a fit to eq. S2-1 for the variation with charge density.

Figure S1 displays the pronounced change in the G⁻ Raman spectra for the different electrostatic gating conditions. The broad background in the spectrum at 0V arises from electronic Raman scattering[S6] and has been subtracted out for the line-width analysis. The inset in Fig. S1 displays the measured gate dependence of the line width of the G⁻ mode (red circles). We fit these data considering the gate-variable electron-phonon coupling[S4, 5], which can be approximated as

$$\Gamma(E_F) = \Gamma_0 + \Gamma_{e-ph}[f(T, E_F, -E_{ph}/2) - f(T, E_F, E_{ph}/2)]. \quad\quad (S2\text{-}1)$$

Here $\Gamma_0$ and $\Gamma_{e-ph}$ are constants, $E_{ph}$ is the energy of the G⁻ phonon, and $f(T, E_F, E)$ is the Fermi distribution function at temperature $T$, where we have neglected the effect of the finite temperature on the chemical potential $\mu$ and have set $\mu = E_F$. In the fit, we assume that the Fermi energy $E_F$ in the metallic nanotube is directly proportional to the gate voltage $V_G$. In filling the linearly dispersing metallic band, this assumption is equivalent to that of a direct proportionality between $V_G$ and the induced charge density in the nanotube, i.e., as expected for a constant gate capacitance.

From the fit above and the relation $\rho = 4E_F/\pi\hbar v_F$ for charge density in a metallic band with Fermi velocity $v_F \approx 10^6$ m/s, we obtain an effective gate capacitance of $C_G = 0.018e/\text{V·nm}$. Given the range of applied voltages and the shift in the Fermi level, we are justified in neglecting the effect of quantum capacitance[S7] and consider $C_G$ to reflect just the geometrical capacitance. We apply this same value for the semiconducting nanotube. The metallic nanotube under study has been assigned as a (18, 3) species. The difference in diameter between this nanotube and our near zigzag semiconducting nanotube employed in the measurements shown in Fig. 2 is only expected to give rise to a difference in geometrical capacitance of ~ 3% with an assumption of its logarithmic dependence on the nanotube diameter.

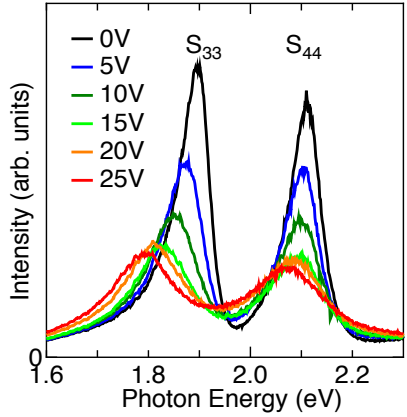

**Figure S2.** Gate-induced changes in the Rayleigh scattering spectra of the near zigzag semiconducting nanotube between $V_G$= 0V to 25V. These changes are similar to those observed for the negative values of $V_G$ that are presented in the main text.

## S3. Rayleigh scattering results for electron doping

Figure S2 shows the Rayleigh scattering spectra of the semiconducting nanotube employed in the measurements described in the main text for the several positive values of the gating voltage $V_G$ corresponding to electron doping. We observed changes similar to those for hole doping shown in Fig. 2a in the main text. The modulation associated with charge injection is thus essentially symmetric for electron and hole doping.

## S4. Evaluation of the effects of strain and electric fields

In our electrostatic gating of a suspended nanotube, the injection of charge leads to electrostatic forces and the generation of axial strain along the nanotube. In order to analyze the effect of strain on the observed peak shifts in the nanotube electronic spectra, we have measured the shift of the Raman 2D (or G') mode simultaneously with Rayleigh scattering measurements. Figure S3 shows the Raman 2D mode spectra of the semiconducting nanotube for various negative gating voltages $V_G$. (Similar changes were observed for positive gate voltages.) The shift of the Raman 2D mode frequency is indicative of applied axial strain[S8]. According to an earlier study[S8], a linear shift with axial strain of ~ 40 cm$^{-1}$/% can be expected for semiconducting carbon nanotubes. Since the maximum value of the observed 2D mode shift is about -30 cm$^{-1}$, the induced strain is inferred to be less than 0.8% at $|V_G|$ = 20V. With this amount of strain, theory[S9] predicts a shift of $S_{33}$ transition energy of about -30 meV, while the observed $S_{33}$ red-shift is about -70 meV at $|V_G|$ = 20V. Thus, while the effect of strain is not dominant for our spectra, it cannot be neglected.

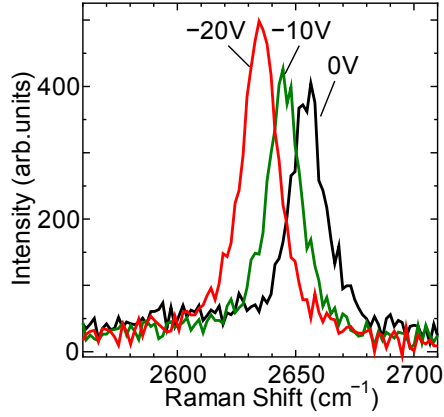

**Figure S3**. Spectra of the Raman 2D mode of the near zigzag semiconducting nanotube for different negative gating voltages $V_G$. The decrease of the Raman 2D (G') mode frequency arises from the presence of uniaxial strain under the electrostatic gating conditions.

Here we discuss how we can essentially eliminate the effect of strain from our measurements and can thus directly investigate the role of many-body corrections. Here we denote the strain-induced peak shift of $S_{ii}$ transition as $\Delta E_{ii}^{S(st)}$ and purely electronic (peak shift from modification of the many-body interactions) contribution as $\Delta E_{ii}^{S(el)}$. The total peak shift $\Delta E_{ii}^{S} = \Delta E_{ii}^{S(st)} + \Delta E_{ii}^{S(el)}$ is the sum of the strain and electronic contributions. According to an earlier study[S9], the strain-induced peak shift for the $S_{33}$ and $S_{44}$ satisfies $\Delta E_{33}^{S(st)} = -\Delta E_{44}^{S(st)}$ to a high degree of accuracy. Thus, the average peak shift of the two transitions,

$$\Delta E_{av} = (\Delta E_{33}^{S} + \Delta E_{44}^{S})/2 = (\Delta E_{33}^{S(el)} + \Delta E_{44}^{S(el)})/2 \quad , \tag{S4-1}$$

can be taken as reflecting only the influence of the *electronic* contributions.

Another possible influence on our experimental measurement is the direct contribution of the applied electric field on the excitonic transition energies, *i.e.*, the potential role of the DC Stark effect in the nanotube. For geometries in which no charging of the nanotube is expected, prior theoretical studies[S10] predicted that a Stark shift on the order of 200 meV for a transverse electric field of 6.7 MV/cm for a zigzag nanotube with a similar diameter to our semiconducting nanotube. In our case, we estimate an effective transverse applied field on the order of 0.01MV/cm, implying a Stark effect of at most 0.3 meV, far less than our observed spectral shifts. In addition, prior experimental investigations[S11] reported the absence of any detectable spectral shifts for transverse electric fields up to 0.2 MV/cm. On this basis, we can ignore the role of the DC Stark effect in our measurements.

**S5. Theoretical treatment of the doping-induced energy shift of the excitonic transitions**

We have analyzed the influence of carrier density on the excitonic transition energies within a *k·p* scheme. The approach involves solving the Bethe-Salpeter (BS) equation[S12, 13]

$$E\psi_k = \left[(E_{c,k} + \Sigma_{c,k}) - (E_{v,k} + \Sigma_{v,k})\right]\psi_k - \sum_q K_{k,k+q}\psi_{k+q}. \tag{S5-1}$$

Here *E* is a complex valued quantity, corresponding to the exciton energy and the line width[S14], *k* and *q* denote wave vectors, $E_{c(v),k}$ is the single-particle band energy, $\Sigma_{c(v),k}$ is the self energy of the conduction (valence) band, and $K_{k,k+q}$ is the Coulomb interaction kernel. In the calculation, the exciton states are determined by a circumferential length $L$ (= $\pi d$) of the nanotube, a characteristic kinetic energy $2\pi\gamma/L$ [$\gamma = (\sqrt{3}/2)a\gamma_0$ with nearest-neighbor hopping integral $\gamma_0$ = 2.7 eV and the lattice constant $a$ = 2.46 Å of graphene], and a characteristic Coulomb energy $e^2/\kappa L$, where $\kappa$ is the effective dielectric constant from the effects of polarization of surrounding materials and electrons far from the Fermi level. The dimensionless Coulomb parameter $v = (e^2/\kappa)/(2\pi\gamma)$ represents the Coulomb interaction $e^2/\kappa L$ scaled by the characteristic kinetic energy. In the calculations, we set $v$ = 0.16, corresponding to $\kappa$ = 2.5.

For calculation of the energy shift, we consider only intravalley scattering near the $K$ ($K'$) valley of the graphene Brillouin zone. We calculate a length-scale ($q$) dependent screened Coulomb interaction $W(q)$ within the static random-phase approximation[S12, 13]. The variation with carrier density is dominated by the $q$-dependent dielectric function $\varepsilon(q)$, which in turn depends on the Fermi energy through the Fermi distribution function. $\varepsilon(q)$ screens the Fourier component of the bare Coulomb interaction $V(q)$ as $W(q) = V(q)/\varepsilon(q)$. The self-energy $\Sigma_{c(v),k}$ is calculated within the screened Hartree-Fock approximation.

We evaluate the dielectric screening induced by the carrier doping within a static random-phase approximation. The dielectric function is expressed as[S12, 13]

$$\varepsilon_{n-m}(q) = 1 + \left[\Pi^{K}_{n-m}(q) + \Pi^{K'}_{n-m}(q)\right]V_{n-m}(q), \quad (S5\text{-}2)$$

where $q$ is the wave vector of the external potential, $n$, $m$ are the band indices that specifies the 1D cutting lines of the nanotube in the graphene Brillouin zone, $\Pi^{K(K')}_{n-m}(q)$ is the polarization function of the $K$ ($K'$) valley, $e$ is the electron charge, and $d$ is the nanotube diameter. $V_{n-m}(q)$ is calculated as

$$V_{n-m}(q) = \left(\frac{2e^2}{\kappa}\right) I_{|n-m|}(|q|d/2) K_{|n-m|}(|q|d/2), \quad (S5\text{-}3)$$

where $I$ and $K$ are modified Bessel functions. $\Pi^{K}_{n-m}(q)$ is calculated from the relation

$$\Pi_{n-m}^{K}(q) = -\frac{2}{A}\sum_{\alpha',\beta',k'}\sum_{k} g_c \delta_{n-m,n'-m'} \left| M_{\beta'k',\alpha'k'+q}^{K} \right|^2$$

$$\times \left[ \frac{f_{\alpha',k'+q}(1-f_{\beta',k'})}{E_{\alpha'}^{K}(k'+q) - E_{\beta'}^{K}(k') + i\delta} - \frac{(1-f_{\alpha',k'+q})f_{\beta',k'}}{E_{\alpha'}^{K}(k'+q) - E_{\beta'}^{K}(k') - i\delta} \right], \quad \text{(S5-4)}$$

where *M* is the scattering matrix element, *A* is the nanotube length, *E(k)* is the energy of corresponding bands, $g_c$ is the energy cut-off function, and $f_{\alpha,k}$ is the Fermi distribution function for $E_{\alpha}^{K}(k)$. A finite imaginary part $\delta/(2\pi\gamma/L) = 0.1$ is introduced in the calculation. The effect of the free-carrier doping is taken into account through the change of the Fermi energy in the Fermi distribution function. The subscripts ($\alpha$, $\beta$) specify the energy bands.

## S6. Theoretical treatment of the doping-induced broadening of the excitonic transitions

Our theoretical treatment for the peak shift within the *k·p* scheme described in section S5 also predicts the exciton line width as the imaginary part of *E* in eq. S5-1. We found, however, that the additional line width is nearly constant for finite carrier densities of $\rho > 0.05$ *e*/nm, and the linear increase of the line width observed in the experiment is not reproduced within this theory. This nearly constant line width can be understood as a consequence of a dominant contribution of intraband, low energy scattering processes involving electrons near the Fermi surface to the calculated line width under sufficiently carrier-doped conditions, where the electronic density of states becomes nearly constant due to hyperbolic band structure of a semiconducting nanotube.

To address this discrepancy, we explored possible contributions of the *intervalley* (*K-K'*) interactions to the line width that were omitted within treatment of the *k·p* approximation, as we now discuss.

The scattering rate of a photoexcited carriers due to the doping in electron-doped nanotubes can be expressed within the second order perturbation theory as

$$\Gamma_e^L = \frac{4\pi A_0^2}{\left(2\pi^2 d\right)^2} \sum_{l_1 l_2 (l_3 = L + l_1 - l_2)} \int dk \int dq \left|M_\nu\right|^2 \delta\left(e_0^L + e_k^{l_1} - e_q^{l_2} - e_{k-q}^{l_3}\right) f_k^{l_1} \left(1 - f_q^{l_2}\right)\left(1 - f_{k-q}^{l_3}\right)$$

$$\Gamma_h^L = \frac{4\pi A_0^2}{\left(2\pi^2 d\right)^2} \sum_{l_1 l_2 (l_3 = L + l_1 - l_2)} \int dk \int dq \left|M_\nu\right|^2 \delta\left(h_0^L + e_k^{l_1} - h_q^{l_2} - e_{k-q}^{l_3}\right) f_k^{l_1} f_q^{l_2} \left(1 - f_{k-q}^{l_3}\right)$$

, (S6-1)

where index $\nu = (k,q,l_1,l_2,l_3)$, $M_\nu$ is the matrix element corresponding to the Coulomb coupling strength[S15], $A_0 = a^3\sqrt{3}/2$, $a$ is the graphene lattice constant, $d$ is the nanotube diameter, and two fold spin degeneracy of the carriers in the first subband has been included. Symbols $e_k^l$ and $h_k^l$ label the absolute energies of the carriers in the conduction and valence bands, respectively, with respect to the nanotube midgap, and $l$ is the angular momentum. The Fermi distribution functions $f_k^l$ depend on the doping level and temperature, here $T = 300$ K. In this analysis of relaxation rates, we neglect the excitonic character of the optical excitations, assuming that the broadening can be modelled adequately in terms of the electron and hole dynamics of the states comprising the exciton.

Within this approximation, Eq. S6-1 implies an additional $S_{33}$ and $S_{44}$ linewidth given by the sum $\Gamma_e^L + \Gamma_h^L$. Because of the electron-hole symmetry in our theoretical treatment, the same scattering rate is obtained in a hole-doped nanotube, but applies equally well for electron-doped nanotubes. The calculated decay rates of the third and forth excited states as a function of doping are shown in Fig. 4c. The agreement with the data is surprisingly good considering that our calculations do not use any adjustable parameters with the value of $\kappa$ (= 2.5) chosen to reproduce the energy shift (Supplemental Information S4). This agreement may be partly fortuitous, due to the cancelation of the excitonic effects not included in evaluating matrix elements by the change of the dielectric response with doping.

The photoexcited electron of the $S_{33}$ state has angular momentum $|L| = 4$, and $e_0^4$ in the $K$ valley decays primarily into a finite $q$-momentum state in the second subband in $K'$ valley ($l_2 = 2$) $e_q^2$ by promoting a first conduction band electron ($l_1 = -1$) $e_k^{-1}$ in the $K'$ valley into the state ($l_3 = 1$) $e_{k-q}^1$ in the $K$ valley with the opposite angular momentum. This scattering process is shown schematically in Fig. 4b and can be summarized as $e_0^4 + e_k^{-1} \rightarrow e_q^2 + e_{k-q}^1$. The excited photohole $h_0^4$ has two decay channels: $e_k^1 + h_q^1 \rightarrow h_0^4 + e_{k-q}^{-2}$ and $e_k^1 + h_q^2 \rightarrow h_0^4 + e_{k-q}^{-1}$ both giving about 66% of the total width of $S_{33}$. The first process, $e_k^1 + h_q^1 \rightarrow h_0^4 + e_{k-q}^{-2}$, in the (26, 0) nanotube contributes only 25% to $\Gamma_h^4$.

The photoexcited electron of the $S_{44}$ state has angular momentum $|L|= 5$, and electron $e_0^5$ in the $K'$ valley decays primarily into a finite $q$-momentum state in the second subband ($l_2$=2) $e_q^2$ by promoting a first conduction band electron ($l_1$=-1) $e_k^{-1}$ in the $K'$ valley into the state ($l_3$=2) $e_{k-q}^2$ in the $K'$ valley with the opposite angular momentum. This process is shown schematically in Fig. 4b and can be described as $e_0^5 + e_k^{-1} \rightarrow e_q^2 + e_{k-q}^2$. The photoexcited hole $h_0^5$ has only one decay channel: $e_k^1 + h_q^2 \rightarrow h_0^5 + e_{k-q}^{-2}$ giving about 40% of the total width of $S_{44}$, which explains the slightly smaller doping induced width observed experimentally for the $S_{44}$ exciton compared to that of the $S_{33}$ exciton.

**Supplementary references**

[S1] L. Huang, X. Cui, B. White, and S. P. O'Brien, J. Phys. Chem. B **108**, 16451 (2004).
[S2] M. Y. Sfeir, T. Beetz, F. Wang, L. Huang, X. M. H. Huang, M. Huang, J. Hone, S. O'Brien, J. A. Misewich, T. F. Heinz, L. Wu, Y. Zhu, and L. E. Brus, Science **312**, 554 (2006).
[S3] S. Berciaud, V. V. Deshpande, R. Caldwell, Y. Miyauchi, C. Voisin, P. Kim, J. Hone, and T. F. Heinz, phys. stat. sol. b **249**, 2436 (2012).
[S4] Y. Wu, J. Maultzsch, E. Knoesel, B. Chandra, M. Huang, M. Y. Sfeir, L. E. Brus, J. Hone, and T. F. Heinz, Phys. Rev. Lett. **99**, 027402 (2007).
[S5] J. C. Tsang, M. Freitag, V. Perebeinos, J. Liu, and P. Avouris, Nature Nanotech. **2**, 725 (2007).
[S6] H. Farhat, S. Berciaud, M. Kalbac, R. Saito, T. F. Heinz, M. S. Dresselhaus, and J. Kong, Phys. Rev. Lett. **107**, 157401 (2011).
[S7] S. Ilani, L. A. K. Donev, M. Kindermann, and P. L. McEuen, Nature Phys. **2**, 687 (2006).
[S8] S. B. Cronin, A. K. Swan, M. S. Ünlü, B. B. Goldberg, M. S. Dresselhaus, and M. Tinkham, Phys. Rev. B **72**, 035425 (2005).

[S9] M. Huang, Y. Wu, B. Chandra, H. Yan, Y. Shan, T. F. Heinz, and J. Hone, Phys. Rev. Lett. **100**, 136803 (2008).

[S10] J. O'Keeffe, C. Wei, and K. Cho, Appl. Phys. Lett. **80**, 676 (2002).

[S11] A. V. Naumov, S. M. Bachilo, D. A. Tsyboulski, and R. B. Weisman, Nano Lett. **8**, 1527 (2008).

[S12] T. Ando, J. Phys. Soc. Jpn. **66**, 1066 (1997).

[S13] Y. Tomio, H. Suzuura, and T. Ando, Phys. Rev. B **85**, 085411 (2012).

[S14] G. Strinati, Phys. Rev. B **29**, 5718 (1984).

[S15] V. Perebeinos, and P. Avouris, Phys. Rev. B **74**, 121410(R) (2006).